\documentstyle[12pt]{article}

\newcommand{\mysection}[1]{\setcounter{equation}{0}\section{#1}}

\newcommand{\nc}{\newcommand}
\nc{\beq}{\begin{equation}} \nc{\eeq}{\end{equation}}
\nc{\beqa}{\begin{eqnarray}} \nc{\eeqa}{\end{eqnarray}}
\nc{\lsim}{\begin{array}{c}\,\sim\vspace{-21pt}\\< \end{array}}
\nc{\gsim}{\begin{array}{c}\sim\vspace{-21pt}\\> \end{array}}
\nc{\k}{K\"ahler$~$}
\nc{\Lagr}{{\cal{L}}}
\nc{\E}{{\cal{E}}}

\begin{document}

\begin{titlepage}
\begin{center}
{\hbox to\hsize{May 1995 \hfill EFI-95-21}}
{\hbox to\hsize{hep-ph/9505244 \hfill JHU-TIPAC-95015}}
{\hbox to\hsize{ \hfill MIT-CTP-2426}}

\bigskip

\bigskip

{\Large \bf
Destabilizing Divergences in Supergravity Theories at Two Loops} \\

\bigskip

\smallskip

{\bf Jonathan Bagger,$^{a,}$\footnotemark[1]
     Erich Poppitz,$^{b,}$\footnotemark[2] }and
{\bf Lisa Randall$^{c,}$\footnotemark[3]}\\

\bigskip

{\small \it $^{\bf a}$ Department of Physics and Astronomy

The Johns Hopkins University

Baltimore, MD 21218, USA }

\bigskip

{\small \it $^{\bf b}$ The Enrico Fermi Institute

University of Chicago

Chicago, IL 60637, USA}

\bigskip

{ \small \it $^{\bf c}$ Center for Theoretical Physics

Laboratory for Nuclear Science and Department of Physics

Massachusetts Institute of Technology

Cambridge, MA 02139, USA}

\bigskip

{\bf Abstract}\\[-0.05in]
\end{center}
We examine the stability of the mass hierarchy in hidden-sector
supergravity theories.  We show that a quadratically divergent
tadpole can appear at two loops, even in minimal supergravity
theories, provided the theory has a gauge- and global-symmetry
singlet with renormalizable couplings to the visible fields.
This tadpole can destabilize the hierarchy.  We also find a
quadratically divergent two-loop contribution to the field-dependent
vacuum energy.  This result casts doubt on the efficacy of
the ``LHC mechanism" for controlling quadratic divergences.  We
carry out the two-loop calculation in a manifestly supersymmetric
formalism, and explain how to apply the formalism in the presence
of supersymmetry breaking to derive radiative corrections to the
supersymmetric and soft supersymmetry-breaking operators.  Our
approach greatly simplifies the calculation and guarantees
consistency of our results with the underlying supergravity
framework.

\bigskip

\footnotetext[1]{Supported in part by NSF grant PHY94-04057.}
\footnotetext[2]{Robert R. McCormick Fellow. Supported in part
by DOE contract DE-FGP2-90ER40560.}
\footnotetext[3]{NSF Young Investigator Award,
Alfred P.~Sloan Foundation Fellowship, DOE Outstanding
Junior Investigator Award. Supported in part
by DOE contract DE-FC02-94ER40818.}

\end{titlepage}

\renewcommand{\thepage}{\arabic{page}}
\setcounter{page}{1}

\baselineskip=18pt

\mysection{Introduction}

The chief motivation for supersymmetry is that it provides the
only field theoretical framework with which to interpret light
fundamental scalars.  For renormalizable theories, this is
due to the fact that supersymmetry prevents quadratically divergent
graphs from renormalizing the scalar potential.  In this sense,
supersymmetry solves the hierarchy problem associated with light
scalar fields.

However, it is expected that a realistic theory will be a
nonrenormalizable
effective field theory, where the nonrenormalizable interactions
are suppressed by a scale $M$.  This might be as large as the
Planck mass, $M_P$, as in supergravity theories, or it might be
some lower scale where other new physics comes into play.  In
either case it is important to understand the extent to which
nonrenormalizable operators affect the hierarchy problem.

When supersymmetry is not broken, the supersymmetric nonrenormalization
theorems ensure that masses are not renormalized in effective
theories.  The nonrenormalization theorems also protect against the
emergence of operators not already present in the superpotential,
even those consistent with gauge and global symmetries.  This leads
to a revised notion of naturalness appropriate for supersymmetric
theories.

When supersymmetry is broken, however, the situation is more subtle.
One would like to know whether radiative corrections induce divergent
operators which destabilize the hierarchy.  One would also like to
know whether supersymmetric naturalness still holds, and in particular,
whether radiative corrections generate all possible soft
supersymmetry-breaking operators consistent with the symmetries.

In essence, supersymmetry reduces by two the degree of divergence
associated with perturbative diagrams.  This is best understood
in superspace, where naive power counting automatically incorporates
supersymmetric cancellations.  The effects of supersymmetry breaking
can be summarized in terms of a (chiral) superspace spurion, $U$,
whose $F$-term contains a supersymmetry-breaking vev.  In hidden-sector
supergravity theories, $U$ is either the conformal compensator
or a hidden-sector chiral field.  In either case, $U = M^2_S
\,\theta\theta$, where $M_S$ is the scale of supersymmetry breaking
\cite{GG}.

In an effective supersymmetric theory, radiative corrections can
induce quadratically divergent contributions to the \k potential.
These include terms of the form
\beq
\delta K\ =
\ {\Lambda^2 \over M^2}\ U^+ U
\ +\ {\Lambda^2 \over M^2}\ U^+N
\ +\ {\Lambda^2 \over M^2}\ N^+U
\ +\ {\Lambda^2 \over M^2}\ N^+N
\ +\ ...\ ,
\eeq
where $\Lambda \simeq M$ is the cutoff and $N$ is an arbitrary
chiral superfield.  For hidden-sector models, where $M = M_P$
and $U = M^2_S\, \theta\theta$, these terms give a contribution
of order $M^4_S$ to the vacuum energy, and generate a tadpole
of order $M^2_S\,N$ in the superpotential.  As we will see, such
terms are dangerous:  they can destabilize the hierarchy and
induce masses of order $M_S$ for the scalar fields.

Recently, there has been a controversy about whether radiative
corrections generate divergent tadpoles in supergravity theories
\cite{BP}, \cite{jain}.
In this paper we resolve the question.  We show, consistent with
the result of Jain \cite{jain}, that {\it at one loop}, there are
no quadratically divergent contributions to singlet tadpoles in
supergravity theories with minimal \k potentials.  We demonstrate,
however, that quadratically divergent tadpoles do appear at two
loops, and that these corrections are sufficient to destabilize the
hierarchy.
Furthermore, we find that tadpoles occur at one loop in models with
nonminimal \k potentials, where the visible sector is directly coupled
to the hidden sector that is responsible for supersymmetry breaking.

The hierarchy can also be destabilized by quadratically divergent
contributions to the (field-dependent) vacuum energy.  In the so-called
no-scale ``LHC models'' of Ferarra, Kounnas and Zwirner \cite{LHC},
the scale of supersymmetry breaking is undetermined at tree level.
Instead, it is fixed by minimizing the one-loop effective
potential.  This requires a careful cancellation of one-loop
quadratic divergences which would drive the scale of
supersymmetry breaking to zero or to the cutoff $\Lambda$.

It might be argued that such destabilizing quadratic divergences
are not important because whatever mechanism cancels the cosmological
constant would presumably eliminate these terms, together with their
potentially destabilizing implications.  However, if the quadratic
divergences could be arranged to cancel, one might hope that
the cosmological constant problem could be addressed
solely in terms of the low-energy effective theory.  In
this case, it might be possible to determine the weak scale
physics by minimizing the low-energy effective potential.

Ref. \cite{LHC} attempts to do precisely this by
ensuring that all quadratic
divergences cancel at one loop.  These divergences are independent
of the superpotential, so they can be arranged to cancel through
scaling relations on the \k potential.
The beauty of this scheme is that the
largest contributions to the cosmological constant automatically
vanish, independent of the details of the weak-scale physics (such
as Yukawa couplings).

In this paper, we use our two-loop calculation to explicitly demonstrate
that this situation does not persist at two loops.  We find a
quadratically divergent two-loop contribution to the vacuum energy
which depends on the Yukawa couplings of the low-energy theory.  It is
difficult to envision a simple field theory mechanism, analogous to that
proposed in \cite{LHC}, through which the quadratic divergence can be
cancelled at two loops and beyond.

The outline of this paper is as follows.  We first review the
reasons for concern about quadratically divergent contributions
to the effective potential.  We identify the dangerous operators
using the counting rules from Appendix A, and then focus on the
problems associated with quadratically divergent contributions
to tadpoles and the (field-dependent) vacuum energy.

In Section 3, we explain why the one-loop contributions to these
quantities may vanish.  We then show that the quadratically divergent
contribution to the tadpole vanishes at one loop in supergravity
theories with minimal \k potentials.
(In Appendix B we extend this result to include higher-derivative
nonrenormalizable operators in the \k potential.)
We also discuss the vanishing of the one-loop vacuum energy in
the LHC models, and explain why
it is crucial for generating and maintaining the hierarchy
in this class of theories \cite{LHC}.

In Section 4, we compute the quadratically divergent two-loop
contribution to the effective potential.  Our technique is explained
in Appendix C.  We use superspace Feynman rules and K\"ahler-Weyl
invariance, which greatly simplifies the calculation and explicitly
maintains the constraints of supersymmetry.  We regulate the divergences
with a simple momentum-space cutoff because it is sufficient to reveal
the most important feature of the calculation, namely that there is
indeed a quadratic divergence, but no term of the form $\Lambda^2 \log
\Lambda$.

In Appendix D, we regularize the theory using higher-derivative operators.
We find a different numerical coefficient for the quadratically divergent
term, but the essential result remains the same:  In theories where
the quadratically divergent one-loop contribution vanishes, the
two-loop contribution is the dominant divergence.

In the conclusion, we discuss the implications
of our two-loop calculation, and
speculate about higher loops.  We argue that higher-loop
quadratic divergences can destroy the hierarchy through
quadratically divergent tadpoles, or by interfering with the
mechanism that underlies the LHC models.  The end result
is that light gauge- and global-symmetry singlets are dangerous,
and that LHC models require additional cancellations beyond
one loop.

\mysection{The Destabilizing Consequences of Quadratic Divergences}

In renormalizable, globally supersymmetric theories, the naturalness
of the hierarchy is guaranteed by the absence of quadratic divergences.
When supersymmetry is softly broken, quadratic divergences
do not appear at one loop and beyond.

In this paper we consider effective supersymmetric theories coupled to
supergravity.  These theories necessarily contain supergravity-induced
nonrenormalizable terms suppressed by powers of $1/M_P$, as well as
other possible nonrenormalizable terms in the \k and superpotentials.
We assume that supersymmetry is spontaneously
broken, at a scale $M_S$, in a
hidden sector which is coupled to the visible sector by interactions
of gravitational strength. In the limit $M_P \rightarrow \infty,$ we
take the gravitino mass $M_{3/2} \simeq M_S^2/M_P$ to be held fixed,
so the theory that describes the visible fields is globally
supersymmetric with explicit soft supersymmetry-breaking terms.

As discussed above, supergravity is nonrenormalizable, so our theory
must be considered as an effective theory, valid below a cutoff
$\Lambda \simeq M_P$.  Below the Planck scale, the lagrangian that
describes the visible sector can contain any nonrenormalizable terms,
consistent with the symmetries, with coefficients suppressed by
appropriate powers of $M_P$.
In general, after supersymmetry breaking, this leads to
supersymmetry-breaking terms which are not soft.
When inserted into quadratically divergent graphs, they may generate
renormalizable terms whose coefficients scale with positive powers of
$\Lambda$.  Absorbing these terms into the low-energy parameters requires
enormous fine-tuning and destabilizes the hierarchy in the visible sector.

In Appendix A, we argue that the dangerous diagrams
are tadpole and vacuum graphs. These operators can
have coefficients that scale with positive powers of $M_P$.
In principle, mass terms can also be dangerous if they
scale as a single power of the soft supersymmetry-breaking
parameter $M_{3/2}$ and a single power of the cutoff.
However, we show that such mass terms are relevant only
in theories with a singlet where some field has a
Planck-scale vev.

\subsection{Tadpoles}

In this section we show that quadratically divergent tadpoles will
destabilize the hierarchy, provided the corresponding fields are
light, and have renormalizable couplings to the other fields in the
visible sector.\footnote{See also \cite{Ellwanger,Krasnikov}.  For
a discussion in the context of globally supersymmetric grand
unification, see \cite{global}.}
For simplicity, we will consider a supersymmetric
theory in which the visible-sector fields include a vector-like Higgs
representation, $H_i$ $(i = 1,2)$, and a gauge singlet, $N$.  This
toy model retains all the essential features for a discussion of
destabilizing divergences.

We will first assume that the \k potential is minimal, in which case
it splits into a sum of two pieces:  one that involves the visible-sector
fields, and the other their hidden-sector counterparts.  This assumption
is probably highly unnatural at the level of the nonrenormalizable terms
in the \k potential.  However, we will first explore the consequences of
nonrenormalizable visible-sector couplings, and only later generalize
our analysis to the case of nonminimal, mixed-\k terms.

In what follows we assume a renormalizable superpotential,
\beq
W ~=~ \lambda  N H_1 H_2 ~,
\label{super}
\eeq
as well as a general visible-sector \k potential
\beq
\label{kahler}
K_{visible} ~=~  N^+ N ~+~ H^{i +} H_i
{}~+~ {\alpha\over M_P} ~(N + N^{+})  ~H^{i +} H_i
{}~+~ {\beta \over M_P}~\left( {\cal D}^{\alpha} H_1 ~{\cal D}_{\alpha} H_2
{}~+~{\rm h.c.} \right) ~+\ ....
\eeq
In eq.~(\ref{kahler}), we include all terms consistent with
symmetry,\footnote{For simplicity we omit any vector superfields
because they are inessential in our analysis.} up to order $1/M_P$.

We imagine that this theory is coupled to supergravity in the standard
manner \cite{WB}.  We also assume the existence of a hidden sector which
is coupled to the observable sector by gravitational interactions.  The
hidden sector is assumed to spontaneously break supersymmetry at a scale
\beq
\label{msusy}
M_{S}^4 ~=~ \langle V_{hidden} \rangle ~=~
\langle ~ K^{ i j*}  ~ D_i W ~ D_{j *} \bar{W}  \rangle ~,
\eeq
where the summation is over the hidden sector fields,
\beq
\label{covder}
D_i W ~\equiv~ \partial_i W ~+~ {1\over M^2_P}\ K_i W
\eeq
is the K\"ahler-covariant derivative of the superpotential,
and $ K^{ i j*}$ is the inverse \k metric.  The scale
$M_{S}$ is of order $10^{11} - 10^{14}$ GeV, depending on the details of
the supersymmetry breaking.

As usual, spontaneous supersymmetry breaking in the hidden sector
leads to explicit soft supersymmetry breaking in the visible sector.
In components, one finds
\beq
\label{soft}
V_{soft} ~=~ M_{3/2}^2 ~( n^*n + h^{i *} h_i ) ~,
\eeq
where $n,\ h_i$ denote the scalar components of the superfields
$N,\ H_i$.  The Planck-suppressed terms in (\ref{kahler}) induce
hard supersymmetry-breaking operators, such as the nonholomorphic
trilinear scalar interactions
\beq
\label{hard}
V_{hard} ~=  ~ {M_{3/2}^2
\over M_P} ~( n^* + n ) ~h^{i *} h_i ~.
\eeq
It is important to note that there is a definite relation between the
coefficients in eqs. (\ref{kahler}), (\ref{soft}) and (\ref{hard}).
This relation is important for the cancellation at the one-loop
order (see Sect. 3); it does not persist to higher loops.

Terms of the type (\ref{hard}) are known to introduce quadratic divergences
\cite{GG}, and can potentially destabilize the hierarchy if they induce a
quadratically divergent tadpole.  To see this, consider the scalar component
$n$ of $N$.  Let us assume that $n$ has mass $M_N$.  This may be a soft
mass, in which case it is of order $M_{3/2}$, or an arbitrary supersymmetric
mass from an $M_N N^2$ term in the superpotential.  If a tadpole is generated,
the scalar potential for $n$ becomes
\beq
\label{npotential}
M_{3/2}^2 ~ {\Lambda^2 \over M_P} ~( n + n^* ) ~+~ M_N^2~ n^* n ~.
\eeq
The vev of $n$ shifts by
\beq
\label{nshift}
\langle \delta n \rangle ~\simeq~
{\Lambda^2 \over M_P} ~ \left( {M_{3/2} \over M_N} \right)^2 ~,
\eeq
in which case the fields $H_i$ acquire a mass of the order
\beq
\label{mhiggs}
\mu_{12} ~\simeq~ {\Lambda^2 \over M_P} ~ \left( {M_{3/2} \over M_N}
\right)^2 ~.
\eeq

For the hierarchy to be stable, we must require $\mu_{12} \simeq M_{3/2}$.
In this case, eq. (\ref{mhiggs}) implies that the cutoff $\Lambda$ must
be less than $\sqrt{M_N^2 M_P/M_{3/2}}$.  Naturalness places an upper
bound on the scale of new physics in hidden-sector theories with
visible singlets.\footnote{As will be evident from
sect. 4, it is also possible to induce a direct $h_1 h_2$-mass term
if there are vevs of order $M_P$ in the hidden sector.}

For the case of a light singlet, with $M_N \simeq M_{3/2}$, the upper
bound becomes $\Lambda \lsim \sqrt{M_{3/2} M_P}$, which is precisely
the scale of supersymmetry breaking in the hidden sector.  If we turn
this around, and let the cutoff be $M_P$, we find that the singlet must
be heavier than the intermediate scale, $M_N \gsim \sqrt{M_{3/2} M_P}$.

Thus we have seen that the hierarchy is destabilized in the presence
of gauge- and global-symmetry singlets with renormalizable visible-sector
interactions -- {\it provided} that a quadratically divergent tadpole
is generated.  In Sect.~4 we will see that such tadpoles do not
always appear at one loop, but are, in fact, induced at two-loop order.

\subsection{Vacuum Energy}

Quadratically divergent contributions to the vacuum energy represent
a serious unsolved problem in relation to the cosmological constant.
They also destabilize the hierarchy in effective supergravity theories
with a sliding gravitino mass \cite{LHC}.  In these models, the gravitino
mass either turns out to be too large (of order the Planck mass) or
too small (zero).  The stability of the hierarchy requires that the
cosmological constant vanish to order $M_{3/2}^4$.

The authors of Ref. \cite{LHC} enumerated a set of conditions under
which the vacuum diagrams cancel to one loop.  They found that the
cancellation requires a relation between the hidden- and visible-sector
contributions to the one-loop effective potential.  Furthermore, they
proposed that the cancellation does not depend on the parameters of the
superpotential and can be understood in terms of a geometric property
of the K\"ahler potential.

If this situation were to persist at higher-loop order, it would
allow the effective superpotential to be determined dynamically, as
in Ref.~\cite{LHC}.  If, however, the higher-order
visible-sector contributions introduce new quadratic divergences
that {\it depend on the superpotential}, the simplicity of this
mechanism would be called into question.  The point is that in
this case, the cancellation would require the hidden sector to
depend on the same Yukawa couplings as the visible sector.  Furthermore,
the hidden-sector fields would have to be sensitive to the same
radiative corrections as the visible-sector fields.  Such a
situation would be unprecedented and would probably require a
miracle of string theory.  Moreover, the fact that the visible
and hidden sectors depend on the same Yukawa couplings also calls
into question the calculability of these theories.

For this reason, we believe that it is important to calculate the
effective potential to two-loop order.  In Sect.~4 we will find a
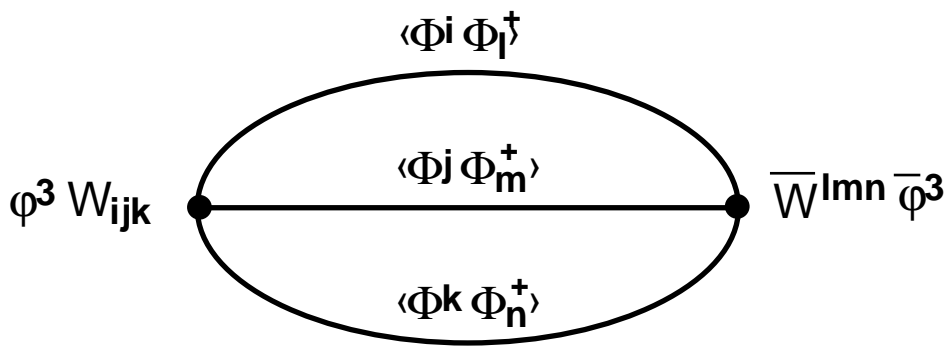
Yukawa-dependent, two-loop, quadratically divergent contribution
to the vacuum energy.\footnote{This contribution to the vacuum
energy was also discussed in \cite{Choi}.}  This contribution will
destabilize the hierarchy in the LHC models.

\mysection{The Dangerous Graphs at One Loop}

\subsection{Tadpoles}

In Ref.~\cite{jain}, it was shown that no quadratically divergent
tadpoles are induced in hidden-sector models with minimal \k
potentials.  This conclusion is based on Refs.~\cite{maryk,marykpv},
in which the divergent parts are calculated for the bosonic
contribution to the one-loop supergravity effective action.
This result is specific to spontaneously broken supergravity
and, as shown below, relies on a cancellation between the mass
and wave function renormalizations.
In this section we will confirm this result in terms of our toy
model.  In the  sect. 4 we shall see that the cancellation is
an accident of the one-loop approximation.

In what follows we will ignore the higher-derivative
term\footnote{In appendix B, we show that the higher-derivative
term with coefficient $\beta$ is equivalent to the trilinear term
with coefficient $\alpha$, to order $1/M_P$.} in (\ref{kahler}).
Therefore the kinetic terms for the matter fields are simply
\cite{WB}
\beq
\label{kinetic}
K_{ij *}\partial_{\mu} z^i \partial^{\mu} z^{j *} ~+~
i K_{ij *} \bar{\chi}^i \sigma^\mu \partial_\mu \chi^{j *}~.
\eeq
In this expression, the \k metric depends on the coefficient
$\alpha$ through field-dependent terms.

In a similar way, the scalar potential is just
\beq
\label{potential}
V ~= ~\exp(K/M_{P}^2)~\left( K^{i j*}~ D_i W~ D_{j*} \bar{W} ~-~
{3 W \bar{W} \over M_{P}^2} \right)~.
\eeq
Here $K = K_{visible} + K_{hidden}$, $W = W_{visible} + W_{hidden}$.
  For our toy model, we cancel the
cosmological constant by adjusting a constant term
\beq
\label{w0}
W_0 = M_{S}^2 M_{P}/\sqrt{3}
\eeq
in the superpotential.

The constant $W_0$ introduces soft supersymmetry-breaking terms in the
scalar potential of the observable sector.  These terms can be found
from (\ref{kahler}) and (\ref{potential}).  Expanding in inverse
powers of $M_P$, we find the following scalar potential for the
visible fields:\footnote{We have assumed that there are no terms in
$K$ which are bilinear both in the hidden and observable fields.
Such terms contribute to nonuniversal soft scalar masses.}
\beq
\label{softlagrangian}
V_{obs}~=~ K^{i j*}~ W_i~  \bar{W}_{j*} ~+~
M_{3/2}^2\  ~K^{i j*} ~ K_i~ K_{j*} ~+~
{\cal{O}}\left( {M_{3/2}^2\over M_{P}^2}
 \right) ~.
\label{totalv}
\eeq
Here $K$ is the \k potential (\ref{kahler}) for the observable fields,
and $M_{3/2}$ is the gravitino mass, $M_{3/2} = M_{S}^2/\sqrt{3} M_{P}$.
The first term in (\ref{softlagrangian}) is the usual supersymmetric
scalar potential, while the second contains the terms that break
supersymmetry.

In terms of component fields, it is not hard to see that the following
terms have the potential to contribute to a quadratically divergent
one-loop tadpole for the singlet field $n$:
\beq
\label{t1}
\alpha~{M_{3/2}^2 \over M_{P}} ~( n + n^*)~ h^{i *} h_i~,
\eeq
from (\ref{totalv}), (\ref{hard}), and
\beq
\label{t2}
{\alpha \over M_{P}}~ n ~\partial_{\mu} h^{i *} \partial^{\mu} h_i ~,
\eeq
which follows from (\ref{kinetic}), (\ref{kahler}).  These terms give rise
to two quadratically divergent one-loop graphs:  one with the supersymmetric
term (\ref{t2}) at the vertex and a soft supersymmetry-breaking insertion
(\ref{soft}) in the propagator, and another with the supersymmetry-breaking
vertex (\ref{t1}) and no insertions in the propagator.  The relative sign
and magnitude of these two contributions are such that the quadratic
divergence cancels.

This confirms Jain's result from Ref. \cite{jain}, in which he
 pointed
out that the kinetic energy and potential renormalizations are such that
all quadratically divergent tadpoles cancel after rescaling the fields to
their conventional normalization. The cancellation in terms
of Feynman graphs represents the same physics.

As this example illustrates, the one-loop cancellation of the singlet
tadpole follows from the assumption of a minimal \k potential in the
supergravity lagrangian.  The cancellation relies on the fact that the
proportionality constant between the soft supersymmetry-breaking mass
term and the quadratic term in the \k potential is precisely the same
as that between the supersymmetry-violating trilinear scalar vertex and
the trilinear term in the \k potential.  Should there exist nonrenormalizable
terms in the \k potential that mix the hidden- and visible-sector fields,
such as $\Phi^+ \Phi H^+ H N/M_{P}^3$ or $\Phi^+ \Phi H^+ H/M_{P}^2$,
where $\Phi$ is a hidden-sector field with nonzero $F_\Phi$, then there
would be additional contributions to the soft supersymmetry breaking that
can destroy the proportionality relations.  A similar conclusion was
reached in \cite{jain}.

This one-loop cancellation can also be understood from a different point
of view.  After the superfield redefinition
\beq
\label{redef1}
H_i ~\rightarrow~ H_i ~\left( 1 ~-~ \alpha ~{N \over M_{P}} \right)\ ,
\eeq
the \k potential (\ref{kahler}) and superpotential (\ref{super})
become (with $\beta = 0$)
\beqa
K ~&=&~ N^{+} N ~+~ H^{i +} H_i ~+~
{\cal{O}}\left({1\over M_{P}^2}\right)~,
\label {redefk}\\
W ~&=&~ \lambda N H_1 H_2 ~-~ \lambda ~\alpha~{N^2 \over M_{P}}~H_1 H_2
{}~+~ {\cal{O}}\left({1\over M_{P}^2}\right)~.
\label{redefw}
\eeqa
After the field redefinition, there are no interactions that could
possibly generate a one-loop quadratically divergent
tadpole.\footnote{This result is reminiscent of the Ademollo-Gatto result,
in which a symmetry is explicitly broken, and the leading correction can be
defined away, but a symmetry-breaking effect appears at next-to-leading
order.}  This field redefinition eliminates both component terms
(\ref{t1}, \ref{t2}) that contribute to the one-loop tadpole.
The field redefinition
also simplifies the two-loop calculation
in sect.~4.

\subsection{Vacuum Energy}

In this section, we will discuss the one-loop contribution to the
(field-dependent) vacuum energy.  As explained
previously, we will work in superspace, and exploit the fact that
all divergent graphs assemble into contributions to the \k potential.
When necessary, we will introduce supersymmetry breaking by inserting
$F$-type vevs into our superspace expressions.

In rigid supersymmetry, the one-loop quadratically divergent contribution
to the \k potential is known to be
\beq
\label{oneloopk}
 {\Lambda^2 \over 16 \pi^2} ~\int d^4 \theta~{\rm log~ det}~ K_{i j*} ~.
\eeq
Supergravity introduces an additional contribution, proportional to
\beq
{\Lambda^2 \over 16 \pi^2} \int d^4 \theta ~{K \over M_{P}^2}~.
\eeq
In this paper, we will not attempt to calculate the precise coefficient
of the supergravity term; this has been done in \cite{superspace}.
Note that the supergravity term does not involve the superpotential.

Both of these terms contribute to the one-loop quadratically divergent
vacuum energy.  After inserting supersymmetry-breaking $F$-term vevs,
one finds contributions of the form
\beqa
\label{vacen}
 {\Lambda^2 \over 16 \pi^2} \int d^4 \theta ~{\rm log~ det} K_{i j*} ~&=&~
 {\Lambda^2 \over 16 \pi^2} ~({\rm log~ det}~ K_{i j*})_{\ell m*}
F^\ell ~ F^{* m} ~,\nonumber\\[3mm]
{\Lambda^2 \over 16 \pi^2} \int d^4 \theta ~{K \over M_{P}^2} ~&=&~
{\Lambda^2 \over 16 \pi^2}  {K_{\ell m*} F^\ell F^{* m} \over M_{P}^2}~.
\eeqa
These terms collect themselves into the one-loop effective potential,
\beq
\label{oneloopv}
V_{1-loop} ~=~ {1\over 32 \pi^2}~ \Lambda^2~{\rm Str}
 \left[ {\cal{M}} (M_{3/2}) \right]^2 ~-~ {1\over 64 \pi^2}~
{\rm Str}~ \left[ {\cal{M}} (M_{3/2}) \right]^4 \log {\Lambda^2 \over
\left[ {\cal{M}} (M_{3/2}) \right]^2} ~,
\eeq
where ${\cal{M}}(M_{3/2})$ is the gravitino-mass-dependent mass matrix.

In a usual field theory calculation, the cutoff dependence would be absorbed
by $\Lambda$-dependent counterterms.  In a calculation from string theory,
the sum of the heavy and light modes should turn the cutoff into a physical
scale related to the Planck scale.  In either case, if the ${\rm Str}
{\cal{M}}^2$ does not vanish, there is a quadratically divergent contribution
to the effective potential.

In models with a sliding gravitino mass, the scale of supersymmetry breaking
is determined by the one-loop effective potential.  If ${\rm Str} {\cal{M}}^2
\ne 0$, the natural scale for the gravitino mass is $M_P$ or zero.
If, however, the \k potential is such that ${\rm Str} {\cal{M}}^2 = 0$,
then the gravitino mass may be stabilized at an exponentially smaller scale.
This is the hope that underlies the LHC mechanism.

The LHC models, however, rely heavily on the precise form of the quadratic
divergences.  It is natural to ask whether the one-loop cancellation persists
to higher-loop order.  We address this question in the following section.

\mysection{The Dangerous Graphs at Two Loops}

In the previous section, we saw that dangerous quadratically divergent
tadpoles vanish at one loop in theories with minimal \k potentials.
We also saw that the quadratically divergent vacuum energy vanishes
in a class of supergravity models, the LHC models.  In each case, the
question of naturalness requires a two-loop analysis.  Therefore in
this section, we compute the quadratically divergent field-dependent
effective potential at two loops.   Our result can be used to find the
quadratically divergent tadpole and vacuum energy.

An important difference between one and two loops is that the one-loop
quadratically divergent effective potential depends only on the \k
potential.  At one loop, the vacuum energy is
independent of the superpotential of the theory.  Diagramatically,
however, it is clear that the vacuum energy depends on the Yukawa couplings
at higher-loop order.  In fact, the two-loop quadratic divergence
is due, in part, to the fact that the one-loop \k potential depends
on the Yukawa couplings.  These Yukawa terms violate the proportionalities
among the soft supersymmetry-breaking terms that are necessary for the
cancellation of the tadpole.  They also violate the scaling properties
of the \k potential that were imposed for successful implementation
of the LHC mechanism.  (Of course, the full two-loop answer is not
obtained by substituting one-loop results into the \k potential, but
requires a true two-loop calculation, which we do here.)

As we already mentioned, we choose to do our calculations in superspace.
This is the simplest way (and in practice, the only way) to calculate
quadratic divergences and ensure that the supersymmetric constraints are
maintained.  It is sufficient to calculate in the supersymmetric theory
because all quadratically divergent graphs, including those that involve
soft supersymmetry breaking, are related to manifestly supersymmetric graphs
by the insertion of supersymmetry-breaking vevs.  As we will see, the full
power of this procedure is realized when one also enforces the constraints
from super-Weyl-\k invariance, which imply definite relations between the
supersymmetric and supersymmetry-breaking contributions to the effective
potential.

In preparation for our two-loop discussion, we will first describe
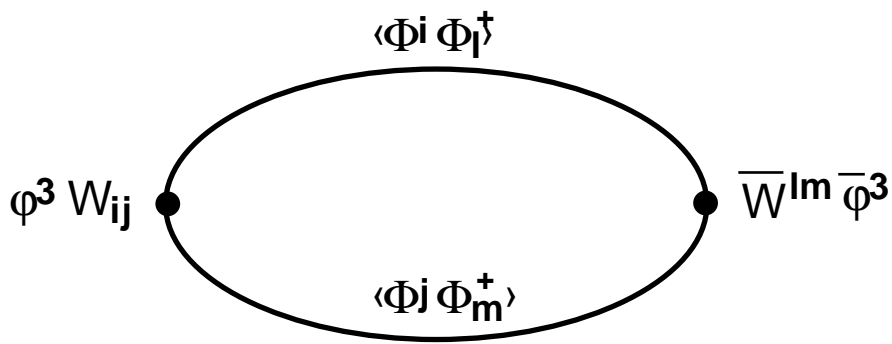
the calculation of the Yukawa-dependent, one-loop logarithmically divergent
contribution to the effective potential.  In the globally supersymmetric
limit, the only logarithmically divergent graph is shown in Fig. 1.  The
calculation is trivial in superspace; it yields the operator
\beq
\label{1loopglobal}
{{\rm log} \Lambda^2 \over 32 \pi^2}
{}~\int d^4 \theta ~ W_{ij} \bar{W}^{ij} ~.
\eeq
When supergravity is included, however, there is much more to the
story.  There are important supergravity corrections to (\ref{1loopglobal}),
which can most easily be found by performing a background-field calculation
in superspace.

In a background field calculation, there appear two types of infinite
counterterms -- those that are invariant under background field
reparametrization (on-shell counterterms) and those that are not
invariant, but vanish on-shell (off-shell counterterms).  The latter
can be eliminated by field redefinitions and do not correspond to
divergences of the S-matrix.  Therefore we are only interested in
the on-shell divergences of the theory.\footnote{For a superspace
discussion of the 2$d$ supersymmetric sigma model, see Ref.~\cite{AGF}.
A related component discussion in 4$d$ supergravity is given in
Ref.~\cite{maryk}.}

In rigid supersymmetry, the expression (\ref{1loopglobal}) is
automatically invariant under redefinitions of the background fields
because the superpotential transforms as a scalar under field
redefinitions.  In supergravity, however, the superpotential is not
an ordinary holomorphic function of the chiral superfields, but is
instead a section of a holomorphic line bundle \cite{BW}.  This means
that under field reparametrizations, both the superpotential and \k
potential transform  under \k transformations
\beqa
\label{transform}
W ~&\rightarrow&~ e^{-F/M_{P}^2} ~W \\
K ~&\rightarrow&~ K ~+~ F ~+~ \bar{F} ~,
\eeqa
where $F$ ($\bar{F}$) is a holomorphic (antiholomorphic) function of
the complex fields.

In the superspace formulation of supergravity, the functions $F$
must be promoted to holomorphic functions of the chiral superfields.
In addition, the \k transformation (\ref{transform}) must be accompanied
by a super-Weyl rescaling of the vielbein (see \cite{WB}, Appendix C, and
eq. (\ref{comptransf}) below).  The resulting super-Weyl-\k transformations
impose important constraints on the superspace theory.  They ensure
that the component action is K\"ahler invariant after eliminating the
auxiliary fields and rescaling the metric to impose a canonical
normalization on the Einstein action.

The on-shell divergent counterterms in spontaneously broken supergravity
must respect the symmetries of the underlying theory and therefore be
super-Weyl-\k invariant.  The invariant counterterm can be obtained
from (\ref{1loopglobal}) by inserting a factor of $e^{2 K/ 3 M_P^2}
\varphi \bar{\varphi}$ in the integrand.  One finds\footnote{Details
of the derivation of (\ref{1loopk}) are given in Appendix C.}
\beq
\label{1loopk}
{{\rm log} \Lambda^2 \over 32 \pi^2} ~\int d^4 \theta
{}~ e^{2K / 3 M_P^2} ~\varphi \bar{\varphi} ~W_{ij} \bar{W}^{ij}~,
\eeq
where $\varphi$ is the conformal compensator superfield of supergravity
theory.  In addition, the derivatives must be K\"ahler-covariant, which
implies
\beqa
\label{covariant}
D_i ~&\equiv &~ e^{K/M_{P}^2} ~\partial_i ~ e^{- K/M_{P}^2}\nonumber \\
W_i ~&\equiv &~ D_i W \nonumber \\
W_{ij} ~&\equiv &~ D_j W_i \ -\ \Gamma_{i j}^k W_k ~,
\eeqa
where $\Gamma_{i j}^k$ is the connection of the \k manifold.

The chiral compensator $\varphi$ transforms as follows under super-Weyl-\k
transformations,
\beqa
\label{comptransf}
\varphi ~&\rightarrow&~ e^{F/3 M_{P}^2} ~\varphi \nonumber ~\\
\bar{\varphi} ~&\rightarrow&~ e^{\bar{F}/3 M_{P}^2} ~\bar{\varphi} ~.
\eeqa
With these transformations, it is easy to see that the term (\ref{1loopk})
is super-Weyl-\k invariant.

This one-loop term in the effective \k potential has important physical
consequences.   The rigid piece of (\ref{1loopk}) determines
 the anomalous
dimensions of the chiral superfields.  The vevs of
the $e^{2 K/3 M^2}$ factor and of the chiral compensators play the
roles of spurions, so (see Appendix C)
\beqa
\label{compvevs}
e^{2 K/3 M^2_P} ~&=&~ e^{2 K/3 M^2_P} ~\left[
{}~1 ~+~ \theta^2 ~{2\over 3} K_i F^i
{}~+~ \bar{\theta}^2~ {2\over 3} K_{i*} F^{* i} \right. \nonumber \\
{}~&&~~~+~ \left. \theta^2 \bar{\theta}^2 ~{2 \over 3} \left( K_{i j*} ~+~
 {2 \over 3} K_i K_{j*} \right)~ F^i F^{* j} ~\right] \nonumber \\[3mm]
\varphi ~&=&~ e^{K / 6 M_{P}^2} \left[ 1 +
\theta^2 \left( e^{K / 2 M_{P}^2} {\bar{W}\over
M_{P}^2} ~+~ {K_i F^i \over 3 M_{P}^2}\right) \right] \\[3mm]
\bar{\varphi} ~&=&~ e^{ K / 6 M_{P}^2}\left[ 1 + \bar{\theta}^2
\left( e^{K / 2 M_{P}^2} {W\over
M_{P}^2} ~+~ {K_{i*} F^{* i} \over 3 M_{P}^2}\right) \right] \nonumber~,
\eeqa
where terms on the l.h.s. should be interpreted as superfield vevs,
while the terms on the r.h.s. are functions of the vevs of the scalar
components of the chiral superfields. Inserting these expressions
into the logarithmically divergent counterterm (\ref{1loopk}), we
find the logarithmically divergent terms
that are responsible for the
running of the soft scalar
masses,
\beqa
\label{softmasses}
{{\rm log} \Lambda^2 \over 32 \pi^2} ~\int d^4 \theta
{}~ e^{2 K/ 3 M_P^2} ~\varphi \bar{\varphi} ~W_{ij} \bar{W}^{ij}
{}~\supset ~ {{\rm log} \Lambda^2 \over 32 \pi^2}\ {M_S^4 \over M_{P}^2}
{}~ \lambda_{ijk} \bar{\lambda}^{ij}{}_{\ell} ~\phi^k \phi^{* \ell}\ .
\eeqa
Here the $\lambda_{ijk}$ are the Yukawa couplings and $\phi^k$ are
the scalar components of the visible-sector fields.
To derive this expression, we also used the facts that
\beq
K_{\ell m*} F^\ell F^{* m}\ =\ M^4_S
\eeq
 and
\beq
{W \bar{W} \over M^4_P}\ =\ {M^4_S \over 3 M^2_P} ~.
\eeq

{}From the expression (\ref{softmasses}), one can immediately find the
Yukawa-coupling-dependent part of the one-loop beta functions for the
soft masses.  Note that one would have found the same answer for the
soft masses by inserting $e^K$ instead of $\varphi \bar{\varphi}
e^{2 K/3}$.  However, this procedure does {\it not} give the correct
answer for the soft supersymmetry-breaking trilinear terms.  This
illustrates the importance of the compensator formalism in superspace
supergravity theories.

This example demonstrates that the leading (in terms of $1/M_P$)
divergent contributions to the visible-sector effective potential
can be found by the following procedure:

\begin{enumerate}
\item
Calculate the divergent part of a rigid supergraph.

\item
Cast the resulting operator in a form that is invariant under supergravity
field redefinitions.  For graphs with only chiral vertices and no purely
chiral propagators, this means multiplying the integrand by a factor of
\beq
\left( e^{K/3 M_P^2} \right)^P ~( \varphi \bar{\varphi} )^{3 V - P}~,
\eeq
where $P$ is the number of $\Phi^+\Phi$ propagators and $V$ the
number of chiral vertices.

\item
Insert supersymmetry-breaking vevs for the chiral compensator
and the hidden-sector fields.
\end{enumerate}

The resulting component expressions give the divergent contributions
to the supersymmetric and soft supersymmetry-breaking operators.
For the case at hand, the result (\ref{1loopk})
coincides exactly with the one obtained by Jain and Gaillard
\cite{maryk} from a component calculation of the one-loop effective
action in an arbitrary bosonic background.

Another effect of supergravity is to induce additional terms that
are suppressed by more powers of $1/M_P$.  These terms are induced by
diagrams with gravitational fields in the loops.  We will not compute
such terms in this paper; at one loop, the logarithmically divergent
terms were found in Ref.~\cite{maryk}:
\beq
\label{additional1loop}
\delta K\ =\ -\ {{\rm log} \Lambda^2 \over 32 \pi^2}~
{}~ e^{K/M_P^2} \left(  2 ~ {W_{i} \bar{W}^{i}\over M_{P}^2}
 + 4~ {W \bar{W}\over M_{P}^4} \right)~.
\eeq
These terms are exactly what one would expect by naive power
counting. Their effects on the renormalization of the low-energy
theory are suppressed by powers of $1/M_P$.

We shall follow the same general procedure in our search for
two-loop quadratically divergent operators.  We will first look
for a quadratically divergent graph in rigid supersymmetry, and
then dress it according to the background field prescription.  In
this way we will find leading two-loop contributions to the
effective potential.

Motivated by LHC models, we will seek a quadratically divergent
graph that depends on the visible-sector superpotential.  (There
are other graphs that depend only on the \k potential.)  Such
a quadratically divergent supergraph is shown in Fig. 2.  After
applying the standard super-Feynman rules, we obtain the following
expression for the operator generated by this graph:
\beq
\label{twoloopkA}
{1 \over 3!}~\int d^4 \theta  ~W_{ijk} \bar{W}^{ijk} ~\int {d^4 k \over
(2\pi)^4}
{}~ {d^4 p \over (2\pi)^4} ~{1\over k^2 q^2 (k + q)^2}.
\eeq
For simplicity, we calculate the integral with a hard momentum-space
cutoff.

The integral in (\ref{twoloopkA}) is clearly quadratically
divergent. Rewriting it as
\beq
\label{integral}
\int\limits_{|k|,|p| < \Lambda} {d^4 k \over (2\pi)^4}
{}~ {d^4 p \over (2\pi)^4} ~{1\over k^2 q^2 (k + q)^2}
{}~=~{\Lambda^2 \over 512 \pi^4} \int\limits_0^1 d x  \int\limits_0^1 d y
{}~{x + y - \vert x -y \vert \over x y}~,
\eeq
where the term on the r.h.s. follows from performing the angular
integration, we see that there are no infrared divergences.  (Note
that the infrared and ultraviolet divergences would be related, since
on dimensional grounds they would appear as $\Lambda^2 {\rm log}
(\Lambda / m)$.)  The absence of the logarithm justifies the
local operator in (\ref{twoloopkA}).
After performing the integral, we find
\beq
\label{twoloopkB}
{1 \over 3!}~ {\Lambda^2 \over 128 \pi^4}~
 \int d^4 \theta  ~W_{ijk} \bar{W}^{ijk}\ .
\eeq
This calculation, with a hard cutoff, shows that there is indeed a
quadratically divergent contribution at two loops. It also shows that
there are no infrared logarithms associated with this graph.

This calculation might be criticized because a hard momentum-space
cutoff violates supersymmetry. In Appendix D, we repeat the calculation
with a higher-derivative regulator \cite{warr}, and still find that
the operator (\ref{twoloopkB}) is induced.  This regulator preserves
supersymmetry provided that the higher derivatives are covariantized
appropriately.  In Ref.~\cite{marykpv}, Pauli-Villars regularization
was used to calculate the quadratically divergent one-loop graphs.
This works only to one loop \cite{warr}; at higher orders it must be
modified by some other scheme, such as higher covariant derivatives.

In the locally supersymmetric case, the operator (\ref{twoloopkB}) can
be written in the same super-Weyl-\k and field-redefinition-invariant
form as (\ref{1loopk}):
\beq
\label{twoloopk}
{1 \over 3!}~{\Lambda^2 \over (16 \pi^2)^2}~\int d^4 \theta
{}~e^{K/M_P^2} ~W_{ijk} \bar{W}^{ijk} ~.
\eeq
Note the absence of a compensator contribution, which follows from the
Feynman rules explained in Appendix C.

The operator (\ref{twoloopk}) has important consequences for the mass
hierarchy.  If we substitute the observable superpotential of our toy
model (\ref{redefw}) into (\ref{twoloopk}), we generate a term of the
form
\beq
\label{2ltadpole}
{\Lambda^2 \over (16 \pi^2)^2}~ 2 \lambda^2 \alpha
\int d^4 \theta ~e^{ K/M_{P}^2} ~{N + N^\dagger \over M_P}\ .
\eeq
This term induces a quadratically divergent tadpole for the scalar
component $n$ of $N$, as can be seen by inserting the hidden-sector
$F$-term vevs into the \k potential.  This tadpole clearly destabilizes
the hierarchy.\footnote{Note (\ref{2ltadpole})
also implies
 that if there are $M_P$ vevs in the hidden
sector, there is also a direct contribution
of order $ M_{3/2} M_P$ to the Higgs mass term
from the $F$-component of the singlet, $F_N \sim h_1 h_2$.}
The operator (\ref{twoloopk}) also induces a field-dependent
quadratically divergent contribution to the vacuum energy:
\beq
\label{2lcc}
V ~=~ -{1 \over 3!}~ {\Lambda^2 \over (16 \pi^2)^2} \int d^4 \theta ~
 e^{K/M_P^2}~ W_{ijk} \bar{W}^{ijk} ~\simeq~-~
{\Lambda^2 \over (16 \pi^2)^2} ~
  {M_S^4 \over M_{P}^2 } ~{1 \over 3!}~ \vert \lambda_{ijk} \vert^2~\ +
\ ...\ ,
\eeq
where the $\lambda_{ijk}$ are the Yukawa couplings in the theory.
As discussed previously, this term will destabilize the hierarchy
of the LHC models.

\mysection{Implications and Conclusions}

In this paper we have studied the question of naturalness in
effective supersymmetric theories.  In particular, we studied
the potentially destabilizing quadratic divergences that are
induced at one- and two-loop order in the effective potential.

We took a careful look at the generation of divergent tadpole
diagrams.  We confirmed the result of Jain, that gauge- and
global-symmetry singlets do {\it not} develop tadpoles at one-loop
order, provided that the \k potential is minimal.  At two loops,
we showed that a quadratically divergent tadpole can indeed be
generated.

Our results indicate that the one-loop result is an accident of
the one-loop approximation.  This conclusion is in accord with
our notions of naturalness because there is no symmetry that
would forbid a divergent tadpole.  Since there is no symmetry,
it has to appear, and indeed it does.

We also took a second look at the so-called LHC models of Ferrara,
Kounnas and Zwirner.  These models rely on a cancellation of the
cosmological constant to order $M_{3/2}^4$.  At one loop such
a cancellation can be enforced by choosing a special \k potential
for the visible and invisible sectors.  Again, since this
is not related to a symmetry of the theory, we expect a
contribution to arise at higher loops. Indeed, at two loops we found
that the cancellation is spoiled by terms that depend on the
superpotential of the visible sector.  Our results imply that
LHC models  do not work unless there is some conspiracy
between the visible and invisible worlds which cancels the
Yukawa-dependent divergence. Such a cancellation would be
difficult to understand at the level of effective field
theory.

Our calculations can be readily generalized to higher loops.  The
superpotential-dependent divergences can be guessed by induction.
At one loop, we found logarithmically divergent contributions to
the component \k potential which go like
\beq
\log(\Lambda^2)~e^{K/M^2_P}\ (W_{ij} \bar{W}^{ij}
{}~+~ {1\over M^2_P} W_i \bar{W}^i
{}~+~ {1\over M^4_P} W \bar{W})\ .
\eeq
At two loops, we found quadratically divergent terms such as
\beq
\Lambda^2~e^{K/M^2_P}\ (W_{ijk} \bar{W}^{ijk}
{}~+~ {1\over M^2_P} W_{ij} \bar{W}^{ij}
{}~+~ {1\over M^4_P} W_i \bar{W}^i
{}~+~ {1\over M^6_P} W \bar{W})\ .
\eeq
Therefore at three loops, we expect quartically divergent
terms of the form
\beq
\Lambda^4~e^{K/M^2_P}\ (W_{ijk\ell} \bar{W}^{ijk\ell}
{}~+~ {1\over M^2_P} W_{ijk} \bar{W}^{ijk}
{}~+~ {1\over M^4_P} W_{ij} \bar{W}^{ij}
{}~+~ {1\over M^6_P} W_i \bar{W}^i
{}~+~ {1\over M^8_P} W \bar{W})\ .
\eeq
The leading term comes from a rigid supersymmetry graph,
while the other terms come from graphs with supergravity
fields in the loops.

Taking $\Lambda \simeq M_P$, we see that the three-loop terms
induce new possibilities for destabilizing divergences.  For
example, the $W_{ijk\ell} \bar{W}^{ijk\ell}$ term also contains
a quadratically divergent tadpole.  Note that this means we
expect new superpotential-dependent divergences at all orders
of perturbation theory.  Moreover, it implies that the
cancellation of quadratic divergences would require a
conspiracy between terms at all orders in the loop expansion.
Clearly, if LHC models are to work,
there must be a deep reason to explain this
miracle.  Presumably, this is related to the cosmological
constant problem, about which (once again)
we have nothing to say.

We would like to thank Mary K.~Gaillard for
discussions about ref \cite{maryk}.
JB and LR would like to thank the Aspen Center for
Physics, where this work was initiated. EP would like to
thank MIT, where this work was completed, for its hospitality.

\newpage

\appendix

\mysection{Superspace Counting Rules}

\bigskip

In an ordinary field theory, operators of dimension three or less
can have coefficients that scale with a positive power of the cutoff.
In a spontaneously broken supersymmetric theory, supersymmetric cancellations
imply that at least one power of the cutoff is replaced by $M_{3/2}$,
the scale of supersymmetry breaking in the visible sector.

The supersymmetric cancellations are such that quadratically divergent
diagrams affect the vacuum energy.  Such divergences can destabilize the
hierarchy in supergravity theories with a sliding gravitino mass \cite{LHC}.
In addition, tadpoles can have quadratically divergent coefficients in
theories with singlets.  Finally, hidden-sector supersymmetric theories
have potentially dangerous mass renormalizations, of order $M_{3/2} M_P
\simeq M_S^2$.  However, divergent mass terms, of order $M_{3/2} M_P$,
can  also  only be generated in theories with singlets.

The divergence structure of spontaneously broken supersymmetric theories
can  be obtained from a manifestly supersymmetric calculation for which
the power counting can be done in superspace.  Superspace power counting
automatically incorporates supersymmetric cancellations, and provides an
efficient way to identify dangerous graphs.  Supersymmetry breaking can be
accommodated by inserting supersymmetry-breaking vevs in the supersymmetric
operators induced by the dangerous diagrams.

The usual formula for the superficial degree of divergence of a $D$-type
superspace diagram can be readily generalized to include nonrenormalizable
operators of the type present in (\ref{kahler}).  It becomes
\beq
 D ~ \le ~ 2 \ -\ E_c\  -\ P_c\ +\ \sum_d d V_d~,
\eeq
where $E_c$ denotes the number of chiral external legs, $P_c$ represents
the number of chiral $\langle \Phi\Phi \rangle$ propagators, and $V_d$
denotes the number of nonrenormalizable operators, suppressed by
$(1/M_{P})^d$.  Using this formula, it is easy to see that a given diagram
is proportional to
\beq
\Lambda^D ~\prod_d~ \bigg({1 \over M_{P}}\bigg)^{d V_d}~ .
\eeq
If we take the cutoff $\Lambda$ to be of order $M_P$, this reduces to
\beq
\label{div}
 D ~ \le ~M_{P}^{2-E_c-P_c}~ .
\eeq

Equation (\ref{div}) implies that a vacuum contribution can be proportional
to $M_{P}^2$, while a superspace tadpole diagram can be at most linearly
divergent, that is, proportional to $M_{P}$.  In a similar way, the divergence
associated with a superspace two-point function can be at most logarithmic,
provided there are no Planck-scale vevs.  Therefore models without singlet
superfields are automatically safe from divergent tadpoles.  In particular,
the minimal supersymmetric standard model is safe from destabilizing
tadpoles \cite{HR}.

\newpage

\mysection{Higher-Derivative Terms and the One-Loop Tadpole}

\bigskip

In this Appendix we will study higher-derivative terms in the \k potential.
To order $1/M_P$, we will find that they are equivalent to trilinear terms,
so they do not induce one-loop quadratically divergent tadpoles.

To be concrete, let us consider the same toy model as in the text,
specified by (\ref{super}), (\ref{kahler}), including the term with
the higher supercovariant derivatives.  After supersymmetry breaking,
this term leads to a Dirac mass for the fermionic components $\chi_i$
of the superfields $H_i$,
\beq
\label{hardfermionmass}
\beta ~{M_{3/2}^2 \over M_{P}} ~\chi_1 \chi_2 ~.
\eeq
The mass (\ref{hardfermionmass}) is hard, and when  inserted into
a one-loop tadpole, it can lead to an uncanceled quadratic divergence
\cite{GG}.  The calculation of Ref.~\cite{maryk} did not include such
higher-derivative terms in the tree-level supergravity Lagrangian.

We shall start our analysis by integrating by parts in the supergravity
Lagrangian,\footnote{We use the identity \cite{zumino} $0 = \int d^4 x d^4
\theta ~\partial_M
(E v^A E_A^M)(-1)^a =  \int d^4 x d^4 \theta ~ E {\cal{D}}_A v^A (-1)^a$.}

\beq
\label{parts}
{\cal D}^{\alpha} H_1 ~{\cal D}_{\alpha} H_2 ~\rightarrow~
{}~-~ H_1 ~{\cal D}^{\alpha} {\cal D}_{\alpha} H_2~.
\eeq
We then perform the field redefinition
\beqa
\label{redef2}
H_1 ~&\rightarrow &~ H_1 ~+~{\beta \over M_P}~
( \bar{{\cal D}}^2  - 8 R ) H_2^{+} \\
H_2 ~&\rightarrow &~ H_2 \nonumber ~.
\eeqa
This field redefinition is manifestly supersymmetric since $(\bar{{\cal
D}}^2 - 8R)$ is a chiral projector \cite{WB,zumino}.

After the field redefinition, the \k potential is simply
\beq
\label{kahler2}
K_{visible} ~=~  N^+ N ~+~ H^{i +} H_i ~+~
\alpha ~{N + N^{+} \over M_{P}}  ~H^{i +} H_i ~+~
 {\cal{O}}\left(
{1\over M_P^2}\right) ~.
\eeq
In this expression, we have neglected terms proportional to $R/M_P \simeq
M_{3/2}/M_{P}$, since after supersymmetry breaking, the vev of the superfield
$R$ is of order $M_{3/2}$ \cite{WB}, and fluctuations around the vev are
suppressed by additional powers of $M_{P}$.

Similarly, the superpotential becomes
\beq
\label{redefw2}
W ~=~ \lambda ~N H_1 H_2 ~+~ {\lambda \beta \over M_{P}}~
N ( \bar{{\cal D}}^2 - 8 R ) H_2^{+} H_2 ~.
\eeq
The second term in $W$ is actually a term in the \k potential
\cite{zumino},
\beq
\label{redefk3}
\int d^2 {\cal{\theta}} {\cal{E}}
N ( \bar{{\cal D}}^2 - 8 R ) H_2^{+} H_2 ~=~
\int d^2 {\cal{\theta}} {\cal{E}} ( \bar{{\cal D}}^2 - 8 R )
N H_2^{+} H_2 ~=~ - 4 \int d^4 {\cal{\theta}} E N H_2^{+} H_2 ~.
\eeq
This shows that the higher-derivative term ${\cal D}^{\alpha} H_1 {\cal
D}_{\alpha} H_2$ in the \k potential (\ref{kahler}) is equivalent to the
trilinear $N H_2^{+} H_2$, to order $1/M_{P}$,

Thus we can conclude that at the one-loop level, the higher-derivative
term does not induce a quadratically divergent tadpole, in contrast
with the naive expectation \cite{BP}.  (We have also verified the
cancellation by explicitly calculating the one-loop component graphs
that are induced by this term.)

\newpage

\mysection{Superspace Loop Calculations in Spontaneously
Broken Supergravity}

\bigskip

As was already mentioned in sect. 2, we assume supergravity to be a
low-energy effective theory, valid below some cutoff scale $\Lambda
\lsim M_{P}$.   We are interested in the predictions of the theory
for the observable-sector interactions.  We assume the validity of
a perturbative loop expansion, under the assumption that visible-sector
fields are defined such that their vevs are less than $M_P$.  (If there
were vevs of order $M_P$, the effective propagator would have to be
generalized to incorporate higher-order quadratic terms.)

For the purposes of our calculations, we restrict our attention to
the leading diagrams that involve Yukawa couplings. (Recall that the
classical \k metric does not depend on the superpotential.)  We work
to two-loop order, so we do not need to consider contributions from
gauge, gravity or hidden-sector loops.\footnote{Gauge and gravity
loops are important for subdominant diagrams and at higher-loop order;
we would need to generalize this analysis to accommodate them.}  In
fact, because we assume $F$-type supersymmetry breaking, we can
neglect gauge interactions altogether.  Furthermore, the hidden-sector
and gravitational fields contribute to the visible-sector operators
only through their vevs.

In this section we present a very powerful method for calculating the
divergent loop contributions to soft supersymmetry-breaking operators,
under these assumptions.  We show how to derive the divergent part of the
soft supersymmetry-breaking effective potential in supergravity
from a calculation in the globally supersymmetric theory. Our methods
permit us to use a  regulator appropriate to the supersymmetric theory,
and furthermore to exploit the power of
the superspace formalism.

Therefore at scales below the cutoff, we consider the visible fields,
plus the gravitational and hidden-sector fields which acquire vevs as
a result of supersymmetry breaking.  The gravitational fields are
contained in an $N =1$ chiral superfield, $\varphi$, known as the
chiral compensator.  The chiral compensator appears in the
superdeterminant of the vielbein through the expansion
\beq
\label{vielbein}
E ~=~ E(H_m)~\varphi \bar{\varphi}~ ,
\eeq
where $H_m$ is the real prepotential superfield which contains
the physical polarizations of the graviton and gravitino.

In superspace, the lagrangian for the relevant part of the coupled
supergravity-matter system is \cite{superspace}, \cite{CO}
\beq
\label{lagrangian}
L ~=~ -3 M_{P}^2 \int d^4 \theta ~\varphi \bar{\varphi}~
e^{-K/3 M_{P}^2}  ~ +
{}~ \left( ~ \int d^2 \theta ~\varphi^3 W ~ + ~ {\rm h.c.} \right) ~,
\eeq
where $K$ and $W$ are the \k potential and superpotential of the hidden-
and observable-sector fields.  Equation (\ref{lagrangian}) is valid
in a flat gravitational background, where $E(H) = 1$, so we assume
that the cosmological constant is canceled by the \k potential (as
in LHC models) or by a constant in the superpotential (\ref{w0}).

In addition to local supersymmetry, this lagrangian possesses a local
super-Weyl-\k symmetry, under which the superpotential, \k potential
and chiral compensator transform as follows,
\beqa
\label{kweyl}
W ~&\rightarrow&~ e^{-F/M_{P}^2}\ W \nonumber \\
K ~&\rightarrow&~ K \ +\ F\ +\ \bar{F} \\
\varphi ~&\rightarrow&~ e^{F/3 M_{P}^2} ~\varphi\nonumber ~\\
\bar{\varphi} ~&\rightarrow&~ e^{\bar{F}/3 M_{P}^2}~\bar{\varphi}
\nonumber~.
\eeqa
This is a symmetry of the ``kinematics" -- the torsion constraints
of $N=1$ supergravity.  It is not, however, a real symmetry of the
lagrangian after the fields have been fixed at their vacuum values,
which are determined by their equations of motion and by the
requirement of
canonical Einstein gravity.  The symmetry is crucial for the
superspace formulation of the theory,  so we explicitly
retain it for all superfield calculations.  It is this symmetry
which leads to the usual component \k invariance, after eliminating
the component auxiliary fields and rescaling the component metric \cite{WB}.

Supersymmetry is spontaneously broken by the vevs of the $F$-components
of the nonpropagating compensator and the physical hidden-sector fields.
These vevs break the super-Weyl-Kahler invariance.
Nevertheless, the compensator formalism permits calculations to be
done in a supersymmetric and super-Weyl-Kahler-invariant manner.  At the
end of the calculation the compensator and the $F$-components of all other
fields are eliminated through their equations of motion.  This simplifies
the calculation because one can use the superspace Feynman rules.
Furthermore, one can find
the supersymmetry-breaking vertices from their supersymmetric
counterparts using the procedure which we now describe.

We will first assume that the lagrangian (\ref{lagrangian}) is the classical
lagrangian of the theory.  We can then determine the tree-level vevs of the
chiral compensator and the hidden-sector fields.  The lowest component of
the chiral compensator is fixed by requiring that (\ref{lagrangian}) yields
the correctly normalized Einstein gravitational action
\beq
\label{einstein}
-3 M_{P}^2 \int d^4 x d^4 \theta\ E(H_m)~=~ -{1\over 2} M_{P}^2
\int d^4 x \,e {\cal{R}} \ +\  ...\ ,
\eeq
where we have restored the part of the supervielbein that contains the
physical graviton and gravitino.  If (\ref{lagrangian}) is to yield
(\ref{einstein}), we must fix super-Weyl-\k gauge by requiring
\beq
\varphi \bar{\varphi}~
e^{-K/3 M_{P}^2}\big\vert_{\theta = \bar{\theta} = 0} ~=~ 1 ~.
\eeq
Fixing this gauge is equivalent to Weyl rescaling the metric in
the component theory.

To find the $F$-components of the fields, we then compute the superfield
equations of motion from (\ref{lagrangian}),
\beqa
\label{compeqn}
-{1\over 4} \bar{D}^2 \left( \bar{\varphi}~
e^{-K/3 M_{P}^2}  \right) ~&=& \varphi^2 W ~\nonumber \\
-{1\over 4} \bar{D}^2 \left( \varphi \bar{\varphi}~
e^{-K/3 M_{P}^2} K_i  \right) ~&=& - \varphi^3 W_i ~.~
\eeqa
Using the transformation law (\ref{kweyl}) it is easy to show that
these equations (\ref{compeqn}) transform covariantly under super-Weyl-\k
transformations.  Taking the lowest component of the first equation in
(\ref{compeqn}), and fixing the lowest component of $\varphi$ by
requiring
\beq
\label{varphisol}
\varphi\big\vert_{\theta = 0}\ =
\ \bar{\varphi}\big\vert_{\bar{\theta} = 0}\ =
\ e^{K/ 6 M_{P}^2}\big\vert_{\theta = \bar{\theta} = 0} ~,
\eeq
we find the following result for the vevs of the chiral compensator
superfields,
\beqa
\label{varphivevs}
\varphi ~&=&~ e^{K / 6 M_{P}^2} \left[ 1 \ +\ \theta^2
\left( e^{K / 2 M_{P}^2} {\bar{W}\over
M_{P}^2} ~+~ {K_i F^i \over 3 M_{P}^2}\right)
\right] \nonumber \\
\bar{\varphi} ~&=&~ e^{ K / 6 M_{P}^2} \left[ 1 \ +\ \bar{\theta}^2
\left( e^{K / 2 M_{P}^2} {W\over
M_{P}^2} ~+~ {K_{i*} F^{* i} \over  3 M_{P}^2}\right) \right] ~.
\eeqa

In a similar fashion, using eqs. (\ref{compeqn}), we can find the
solution for the matter $F$-terms,
\beqa
\label{fvevs}
F^i ~&=&~ - e^{K/2 M_{P}^2} K^{ij*} D_{j*} \bar{W} ~\equiv~
 - e^{K/2 M_{P}^2} \bar{W}^i \nonumber ~,\\
F^{* j} ~&=&~ - e^{K/2 M_{P}^2} K^{ij*} D_{i} W ~\equiv~
 - e^{K/2 M_{P}^2} W^{j*}~.
\eeqa
These tree-level vevs can be inserted into one-loop
graphs to determine the quadratically divergent component operators.
They must be corrected at higher-loop order.

The choice (\ref{varphisol}) explicitly breaks the component \k invariance,
and indeed, the solutions (\ref{varphivevs}) transform noncovariantly.
However, the final component action is still K\"ahler-invariant
because it is obtained from a super-Weyl-\k invariant expression.  This
can be verified explicitly by substituting (\ref{varphivevs}), (\ref{fvevs})
into the lagrangian (\ref{lagrangian}).  We have checked that this procedure
gives the standard K\"ahler-invariant formula for the supergravity scalar
potential (\ref{potential}).

To compute the loop corrections to the lagrangian (\ref{lagrangian})
in the classical background (\ref{varphivevs}), we expand the matter
superfields to second order in the fluctuations $\Phi$ around their vevs:
\beqa
\label{expansion}
L ~=~ L_{class} ~&+& ~\int d^4 \theta ~
\left( \varphi \bar{\varphi} ~e^{- K/3 M_{P}}~ K_{i j*} \right)_{class}
{}~\Phi^i \Phi^{+ ~j} ~+~ ...  \nonumber \\
&+&~ \left( ~\int d^2 \theta ~\left( {1\over 2}
{}~\varphi^3 ~W_{i j} \right)_{class}
\Phi^i \Phi^{ j} ~+~ {\rm h.c.} \right)~.
\eeqa
In this expression, the dots denote terms suppressed by additional powers
of $M_{P}$.  For notational simplicity, we assume that all vevs are
smaller\footnote{One might worry that the
expansion (\ref{expansion}) is not explicitly super-Weyl-\k invariant
since the derivatives of $W$ are not \k covariant. However, the expansion
(\ref{expansion}) is gauge equivalent to a reorganized expansion, obtained
by redefining the chiral compensator: $\varphi = \varphi^\prime \exp{( -
\log{W}/3 )}$.
The whole expansion then only depends on the super-Weyl-\k
invariant $G = K + \log{W \bar{W}}$ \cite{CO}, and therefore manifestly
preserves the super-Weyl-\k invariance.  The Feynman rules, however, are
more cumbersome and for simplicity we prefer to use the expansion
(\ref{expansion}).  We have explicitly checked that the divergent contribution
to the effective action obtained by using this expansion is equivalent to ours,
after using the superfield equations of motion for the background
superfields.} than $M_{P}$.

The Feynman rules that follow from this expansion are easily obtained from
the flat-space super-Feynman rules. The chiral vertices carry an additional
factor of $\varphi^3$, and the chiral field propagators are of the form
\beq
\label{propagator}
\langle \Phi^i (p, \theta) \Phi^{+ ~j}
(p, \theta^\prime) \rangle ~\simeq~
\left( { e^{ K/3 M_{P}} \over \varphi \bar{\varphi}}~
K^{i j*} \right)_{class}
{}~{\delta^4 (\theta - \theta^\prime) \over p^2 }~.
\eeq
Note that the above expression is not the exact propagator in the superfield
background \cite{GRS}.  We have omitted terms where the supercovariant
derivatives act on the background superfields.  These terms reduce the
degree of divergence and do not contribute to the leading divergences
\cite{2loopRGE}.

After applying the above super-Feynman rules, the logarithmically divergent
part of the graph from Fig.~1 is easily seen to be
\beq
{{\rm log} \Lambda^2 \over 32 \pi^2} ~\int d^4 \theta
{}~ e^{2K / 3 M_P^2} ~\varphi \bar{\varphi} ~W_{ij} K^{i l*}
 K^{j m*} \bar{W}_{l* m*} ~.
\eeq
This precisely equals (\ref{1loopk}).  The same Feynman rules, when
applied to the quadratically divergent two-loop supergraph of Fig.~2,
yield the result (\ref{twoloopk}) from Sect.~4.

\newpage

\mysection{The Two-Loop Diagram with a Higher-Derivative
Regulator}

In this Appendix we describe the calculation of the quadratically divergent
supergraph of Fig.~2, using a supersymmetry-preserving higher-derivative
regulator.  As explained in the previous Appendix, we can reduce the
calculation to that of a rigid supersymmetric theory in the background of
the hidden-sector and chiral-compensator fields.  Hence it is sufficient
to employ a rigid version of the higher-derivative regulator. (For a recent
discussion, see Ref.~\cite{kl}).

To regulate the ultraviolet behavior of the chiral-field propagators, we
add to their lagrangian the following term \cite{warr}:
\beq
\label{higherderiv}
L_{regul} ~=~ ~\int d^4 \theta ~
\left( \varphi \bar{\varphi} ~e^{- K/3 M_{P}}~ K_{i j*} \right)_{class}
{}~\Phi^i  f( {\Box\over \Lambda^2} ) \Phi^{+ ~j} ~.
\eeq
The chiral propagator then becomes:
\beq
\label{newpropagator}
\langle \Phi^i (p, \theta) \Phi^{+ ~j}
(p, \theta^\prime) \rangle ~=~
\left( { e^{ K/3 M_{P}} \over \varphi \bar{\varphi}}~
K^{i j*} \right)_{class}~{\delta^4 (\theta - \theta^\prime) \over p^2
( 1 + f(p^2 / \Lambda^2 ))}~.
\eeq
In a theory without gauge fields, this procedure is sufficient to regulate
all divergences \cite{warr}.

To regulate the two-loop graph from Fig. 2, it suffices to take $f(x) = x$,
as we show below.  With this choice the integral (\ref{integral}) becomes
\beqa
\label{newintegral}
I ~&=&~\int {d^4 k \over (2\pi)^4}
{}~ {d^4 p \over (2\pi)^4} ~{1\over (k^2 + {k^4/\Lambda^2}) ~
(q^2 + {q^4 /\Lambda^2})
{}~((k + q)^2 + { ( k+q )^4 / \Lambda^2} )} \nonumber \\[3mm]
&=&~ {\Lambda^2 \over 512 \pi^4} ~\int\limits_0^\infty
d x ~\int\limits_0^\infty d y
{}~ {\sqrt{(1 + x + y)^2 - 4 x y} - |x - y| - 1 \over
(x^2 + x) (y^2 + y)} ~.
\eeqa

The infrared structure of this integral is clearly the same as that of the
integral (\ref{integral}) with the hard cutoff, so (\ref{newintegral}) also
does not contain logarithmic terms.  To investigate the ultraviolet
behavior, it is convenient to reduce $I$ to a one-dimensional integral
\beq
\label{1dint}
I =  {\Lambda^2 \over 512 \pi^4} ~\int\limits_0^\infty d x~
{Q(x) \over x^2 + x} ~,
\eeq
where
\beqa
\label{qofx}
Q(x) ~=~ && -2\,x\,\log (x) + 4\,\log (1 + x) + 4\,x\,\log (1 + x)
\nonumber \\[3mm]
&&+ {\sqrt{4\,x + {x^2}}}\,\log ({{-x + {\sqrt{4\,x + {x^2}}}}\over
       {3\,x + {x^2} + \left( 1 + x \right) \,{\sqrt{4\,x + {x^2}}}}})~.
\eeqa
At infinity, the integrand in (\ref{1dint}) behaves as $ x^{-3} \log x$, and
at zero it is $\sim \log x$.
Therefore the integral converges in the UV and the IR, so it can be taken
numerically.  The result is
\beq
\label{result}
I ~=~ 2.84 {\Lambda^2 \over 512 \pi^4} ~.
\eeq

\end{document}